# Unravelling the signatures of effective spin ½ moments in CeVO$_4$: Magnetization and Heat Capacity study

Dheeraj Ranaut and K. Mukherjee

School of Basic Sciences, Indian Institute of Technology Mandi, Mandi 175005, Himachal Pradesh, India

**Abstract**

The realization of an effective spin ($J_{eff}$) ½ state at low temperatures offers a platform to study the enthralling physics behind the disordered states in certain systems. Here, we report the signatures of magnetic ground state associated with $J_{eff}$ = ½ in CeVO$_4$. Our studies confirm the absence of any ordering or freezing down to 1.8 K. In the low temperature region, the Curie-Weiss fit of the inverse DC susceptibility indicate towards the presence of antiferromagnetic correlations among the Ce$^{3+}$ spins. The calculated value of effective moment (~1.16 $\mu_B$) corresponds to $J$ = ½ with $g_J$ ~ 1.20. Further, the field dependent magnetization curve at 2 K follows a behaviour corresponding to $J$ = ½ Brillouin function with $g_J$ ~ 1.13. Magnetic field dependent heat capacity fits very well with two-level Schottky scheme. Our investigations suggest that CeVO$_4$ can be a promising candidate to realise $J_{eff}$ = ½ properties among 3D spin systems.

## 1. Introduction:

Materials exhibiting effective spin ($J_{eff}$) ½ behaviour at low temperatures have been found to be promising candidates to study the mesmerizing physics behind the dynamically disordered states due to inherent strong quantum fluctuations. In geometrically frustrated magnetic systems, the competing interactions between spins enhance the quantum fluctuations resulting in unusual ground states. Particularly, in spin ($S$) = ½ 2D triangular spin systems, the quantum fluctuations are more robust, which evades the long-range ordering (LRO) even at 0 K. Such systems are proposed to be potential candidates to achieve the highly entangled quantum spin liquid (QSL) states [1-3]. Initial study on $S$ = ½ systems was mainly focussed on 3d ($Cu^{2+}$) based compounds. Recently, this interest has extended to 4$d$, 5$d$ and 4$f$ rare earth based magnetic materials [4-18]. The signatures of $J_{eff}$ = ½ at low temperatures has been observed in various $Ru^{3+}$, $Ir^{4+}$ and $Yb^{3+}$ based compounds, through various experimental as well as theoretical probes. The origin of emergence of this exotic $J_{eff}$ = ½ features has been attributed to combined effects of strong spin orbit coupling (SOC) and crystal field effects. Examples of this kind of systems include 4$d$ honeycomb lattice α-$RuCl_3$ [4], 5$d$ honeycomb lattice $A_2IrO_3$ (A = Li, Na, Cu) [5-7] and $H_3LiIr_2O_6$ [8]. As far as the 4$f$ rare earth compounds are considered, $Yb^{3+}$-based triangular lattices are the best to understand the physics of $J_{eff}$ = ½. These Yb-based compounds include dichalcogenide delafossites with formula $AYbX_2$ (A = Li, Na, Ag; X = O, S, Se) [9-14], $YbMgGaO_4$ [15, 16] and $YbZnGaO_4$ [17]. Along with this, more recently, another $Yb^{3+}$- based compound $K_3Yb(VO_4)_2$, belonging to the family of glaserite, have been found to be a potential candidate to achieve $J_{eff}$ = ½ state [18]. In this system, based on magnetic and specific heat measurements, it is observed that the $Yb^{3+}$- site hosts $J_{eff}$ = ½ moments. As both $Yb^{3+}$ and $Ce^{3+}$ based systems correspond to $S$ = ½, this motivated us to investigate a $Ce^{3+}$ based compound, $CeVO_4$ in same context.

$CeVO_4$ belongs to the family of rare earth orthovanadates with chemical formula $RVO_4$. The compounds of this series show interesting magnetic and optical properties due to the presence of indirect super-exchange magnetic interaction and 4$f$ electron-phonon coupling [19-23]. Recently, this series had gained a lot of interest as some of the compounds of this series exhibits exotic magnetic properties. In $TmVO_4$, magnetic field tuned ferroquadrupolar quantum phase transition is reported [24, 25] whereas, a field tuned quantum criticality has been observed in $DyVO_4$ [26]. Along with this, another member of this series $HoVO_4$ is reported to show signatures of novel QSL state [27]. $CeVO_4$ is an interesting compound due to its $f$-electron structure and complexities arising in its synthesis; as Ce can exist in both $Ce^{3+}$ and $Ce^{4+}$ states. It has been reported to be a potential candidate in the fields of solid oxide fuel anode cells [28], luminescent materials [29] and gas sensors [30]. But, the low temperature magnetic properties of $CeVO_4$ had not been extensively studied. Reports of first principal studies suggest the presence of insulating antiferromagnetic (AFM) ground state in $CeVO_4$ [31, 32]. The ground state term symbol for $Ce^{3+}$ is $^2F_{5/2}$ and the six-fold degeneracy of $Ce^{3+}$ ions is expected to be lifted into three Kramers doublets due to the combined effect of strong SOC and crystal electric field. Hence, it will be interesting to investigate the low temperature magnetic ground state of this compound.

In this manuscript, we report the signatures of magnetic ground state associated with $J_{eff}$ = ½ in $CeVO_4$ at low temperatures through temperature ($T$) and magnetic field ($H$) dependent magnetic and heat capacity ($C$) measurements. Powder X-ray diffraction and X-ray photoemission analysis confirms the presence of single-phase tetragonal structure. $T$ dependent magnetic and $C$ measurements at zero field do not show any signs of long-range ordering as well as spin freezing. In the low-$T$ region, the Curie-Weiss (CW) fit to the DC magnetic susceptibility ($\chi_{DC}$) gives the effective moment corresponding to total angular momentum, $J$ = ½, which is further confirmed by the Brillouin function fitting of field dependent magnetization curve at 2 K, corresponding to $J$ = ½. The $H$ dependent heat capacity fits very well with two-level Schottky scheme. The magnetization and $C$ analysis confirms that $Ce^{3+}$ ions host the $J_{eff}$ = ½ states at low temperatures and thus, our system can be a promising candidate to study the physics behind low temperature magnetic ground states in other 3D spin system.

2. **Experimental Details:**

Polycrystalline sample of $CeVO_4$ was synthesised by the conventional solid state reaction method by using high purity $CeO_2$ and $VO_2$ from Sigma Aldrich. The initial reagents were taken in the stochiometric ratio, grinded and given first heat treatment at 800° C, followed by final sintering at 900° C. Room temperature powder X-ray diffraction (XRD) was performed in the range (10°-90°) using Rigaku diffraction with Cu Kα (λ = 1.54) radiation. The crystal structure was refined by the Rietveld method using FullProf Suite software. X-ray photoelectron spectroscopy (XPS) were obtained using a monochromatic Al Kα (1486.6 eV) X-ray source with an energy resolution of 400 meV and Scienta analyser (R3000). The sample surface was cleaned in situ by scraping with a diamond file until sample has the minimum O 1s signal as measured by Al Kα x-ray. The binding energy was calibrated by measuring the Fermi energy of Ag in electrical contact with the sample. The base pressure during the measurement was $5\times 10^{-10}$ mbar. *H* and *T* dependent magnetic measurements in the temperature range 1.8-300 K were performed using the Magnetic Property Measurement System (MPMS) from Quantum Design, USA. Physical Property Measurement System (PPMS), Quantum Design, USA was used to measure *T* dependent *C* at different fields up to 70 kOe.

## 3. Results and Discussion:

### 3.1 Structural analysis

Rietveld refinements of the XRD data obtained at 300 K shows that $CeVO_4$ crystallizes in the tetragonal structure with the space group $I4_1/amd$ and is in single phase (Fig. 1 (a)). The refinement parameters are $R_p$ = 12.8 %, $R_{wp}$ = 13.9 % and $R_{exp}$ = 10.21 %, with a goodness of fit (GOF) ~ 2.6. The obtained lattice parameters a = b = (7.348 ± 0.006) Å, c = (6.454 ± 0.003) Å and V = 355.62 Å$^3$ are found to be in good agreement with previous reports [33]. The position coordinates and occupancy, obtained from the Rietveld refinement is given in the Table 1. The structure of $CeVO_4$ is comprised of isolated $VO_4$ tetrahedron (orange color) which share corners and edges with $CeO_8$ bisdisphenoids (grey color) (Fig. 1 (b)). Each Ce atom is bonded to four different O atoms at two different distances in form of compressed and elongated bonds (i.e., Ce-$O_c$ and Ce-$O_e$ bonds). Thus, the $CeO_8$ polyhedron is considered to be comprised of two interpenetrating $CeO_4$ tetrahedra. The calculated different bond lengths and bond angles, along with the number of bonds are listed in the Table 1. Further, in order to do a comparative study, we have made a table (Table 2) describing the structural parameters of the different members of the lanthanide series (Ce – Lu) [33]. From here, it can be inferred that the lattice parameters and volume show a continuous decrement as the atomic number increases.

In order to get an insight on the charge states of $CeVO_4$, we have performed XPS at room temperature. Fig. 2 (a, b, c) depicts the Ce 3d, V 2p and O 1s core level XPS spectra. The in-elastic background is subtracted using the Touggard function [34] and the peaks are fitted using the Voigt function. The solid red and green lines in the figure represent the resultant fit and individual fitted peaks, respectively. Fig. 2 (a) reveals the XPS spectra of Ce 3d. Two peaks at 881.3 and 885.5 eV demonstrates the presence of Ce $3d_{5/2}$, while another two peaks appearing at 899.7 eV and 904.2 eV corresponds to Ce $3d_{3/2}$ peaks. The binding energies at 885.8 eV and 904.4 eV represents the $3d_{10}$ $4f_1$ electronic state, corresponding to $Ce^{3+}$ oxidation state [35]. For $Ce^{4+}$ electronic state, the XPS spectrum shows the peaks at 884.82 eV and 882.43 eV binding energies [36]. In the analysed XPS spectra, no peaks corresponding to the above-mentioned energies is found and this rules out the presence of $Ce^{4+}$ oxidation state. The four deconvoluted peaks, shown by the solid green lines, confirm the presence of $Ce^{3+}$ electronic state. In addition, two different peaks of less intensity (shown by the black arrows) belong to vibrational satellites of Ce 3d state. The fitted core XPS spectra of V 2p are represented in the Fig. 2 (b). The V 2p spectrum is characterised by two states: $2p_{3/2}$ and $2p_{1/2}$. The peaks at 516.7 and 524.2 eV are attributed to the dominant $V^{5+}$ oxidation state [37]. Along with this, a very minor peak around 514.8 eV can be seen which is ascribed to the $2p_{3/2}$ trace of the $V^{3+}$ state present at the surface of the sample [38]. This $V^{3+}$ state is believed to reduce to $V^{5+}$ state on exposing to oxygen in the air [38]. Further, Fig. 2 (c) indicates that the O 1s spectra appear as two broad peaks centred around 529.8 eV (major peak) and 531.4 eV (minor peak). The former value corresponds to the lattice oxygen (i.e., $O^{2-}$ of the $CeVO_4$) [39] and the latter one is ascribed to the surface adsorbed oxygen attributed to the defects or off-balance oxygen stoichiometry [40]. Also, in order to investigate the chemical composition of the sample, we have performed energy dispersive x-ray spectroscopy (EDS). The EDS spectrum (not shown here) illustrates the presence of Ce, V and O elements and no other impurities are seen. The weight percentage is found to be consistent with previous results on pure $CeVO_4$ [41].

### 3.2 Magnetic properties

$T$ dependent DC magnetic susceptibility ($\chi_{DC}$) measurements at different applied fields (up to 60 kOe) are performed under the zero-field cooling (ZFC) condition and are shown in the main panels of Fig. 3. As noted from Fig. 3 (a), the 100 Oe and 10 kOe curves follow same path. An inflection point is observed around 160 K for the 100 Oe curve (shown by the black arrow). For better visualization of this behaviour, we have plotted the same in an expanded form (inset of Fig. 3 (a)). It is also to be noted that the curve below and above this inflection point follows the paramagnetic (PM) behaviour. This implies that around 160 K, the Ce spins try to order and with decrease in $T$, the spins again attain the random configuration. To investigate the effect of magnetic field on the inflection point, we have plotted the $T$ dependence of differential susceptibility ($d\chi_{DC}/dT$) at various applied fields, as shown in the inset of Fig. 3 (b). The curve at 100 Oe shows a dip around 162 K which vanishes at 500 Oe. To shed some more light on this observed feature, we have performed temperature dependent AC susceptibility and heat capacity measurements. Fig. 4 depicts the real part of the AC susceptibility ($\chi'$) in the temperature regime of 1.8 – 200 K at various frequencies (31 – 831 Hz) with AC amplitude 2 Oe and zero DC field. The curve does not show any signature of inflection around 160 K. The behaviour can be more clearly visualised from the inset of Fig. 4 which shows the same data in an expanded form. Additionally, the curve does not exhibit any frequency dependent peak, thereby, ruling out the possibility of any kind of spin freezing. The zero-field heat capacity data (discussed in details in the next section) do not show any peak around the inflection point which indicates towards the absence of any kind of long ranged magnetic ordering. Thus, it is believed that around 160 K, the Ce spins try to order in a preferred direction, but the magnetic interactions are not strong enough to induce a long-range magnetic ordering. Below 160 K, the spins again acquire random configuration. Similar kind of behaviour was also reported in a pyrochlore $Na_3Co(CO_3)_2Cl$ where the magnetic susceptibility exhibits an inflection point at 17 K, but the neutron diffraction data do not reveal any long range ordering [42]. Further, as mentioned above, we have performed $T$ dependent $\chi_{DC}$ measurements at various applied fields. Fig. 3 (b) illustrates the $T$ response of $\chi_{DC}$ at $H \geq 30$ kOe. From here, it is evident that $\chi_{DC}$ does not show signs of saturation even at an applied field of 60 kOe, implying that the spins do not tend to align along the direction of applied field. This signifies that the Ce spins prefer to remain disordered even in the presence of high external magnetic field of 60 kOe.

Fig. 5 (a) shows the inverse DC magnetic susceptibility ($\chi_{DC}^{-1}$) vs. $T$ curve at 100 Oe. In the high-$T$ (*HT*) region (above 160 K), the data is fitted with CW law: $(T - \theta_{HT})/C_{HT}$ (shown by the solid red line). The obtained values of the *HT* Curie constant ($C_{HT}$) and *HT* Curie-Weiss temperature ($\theta_{HT}$) are (0.855 $\pm$ 0.001) emu/mol-Oe-K and (–119.631 $\pm$ 0.422) K, respectively. The obtained value matches well with the theoretical value calculated by using the expression given as: $C = N_A(g_J\mu_B)^2 J(J+1)/3k_B$ (where $N_A$, $\mu_B$ and $k_B$ are Avogadro's number, Bohr magneton and Boltzmann constant respectively. $g_J$ and $J$ corresponds to Lande's factor and total angular momentum, respectively), with $J = 7/2$ and $g_J = 0.877$. The calculated value of $g_J$ matches very well with the theoretical value of $g_J = 6/7 \sim 0.857$ for $Ce^{3+}$ state. Also, the experimentally calculated effective magnetic moment $\mu_{eff}$ (= $2.8\sqrt{C}$ $\mu_B$ $\sim 2.58$ $\mu_B$), matches very well with the theoretically calculated value of $\mu_{eff} = 2.54$ $\mu_B$ for $Ce^{3+}$, having $^2F_{5/2}$ multiplet with $g_J = 6/7$. This implies that the thermal energy is comparable to the crystal field splitting in this temperature range and all the states are equally populated. The negative sign of $\theta_{HT}$ signifies the dominance of antiferromagnetic (AFM) correlations among the $Ce^{3+}$ spins. Also, in the *HT* regime, the large value of $\theta_{HT}$ can be attributed to the dominance of crystal electric field (CEF) excitation of $Ce^{3+}$, along with the antiferromagnetic correlations. From the ground state term symbol $^2F_{5/2}$, the total angular momentum is $J = 5/2$ and it exhibits $2J + 1 =$ six-fold degeneracy. In the CEF environment of $D_{2d}$ symmetry, the energy levels split into three Kramer doublets. The ground state doublet is separated from the first excited doublet by an energy difference which is termed as the CEF gap. In many Yb-based systems ($S = ½$), this gap has been calculated and is found to be proportional to the parameter $\theta_{HT}$ [43, 44, and 45]. In these systems, the observed large value of $\theta_{HT}$ is correlated to a large CEF gap. Thus, in our system $CeVO_4$, as $\theta_{HT}$ is large (~ - 110.23 K), the CEF gap is also expected to be large. At temperatures lower than the gap (where the CEF excitations are not that much dominating) a slope change in $\chi_{DC}^{-1}$ is observed. This deviation from the CW law possibly indicates towards the evolution of a ground state CEF doublet with small exchange couplings. Thus, in order to get an insight on the low-$T$ ($LT$) magnetic ground state, we have further plotted the $\chi_{DC}^{-1}$ as a function of $T$ below 20 K and at 100 Oe (Fig. 5 (b)). This data is fitted with the modified CW law: $1/[\chi_0 + C_{LT}/(T − \theta_{LT})]$, as shown by the solid red line [18]. Here, the $\chi_0$ term is composed of the contribution from the core diamagnetic susceptibility and temperature independent Van Vleck paramagnetic susceptibility. This term is included as it contributes to $\chi_{DC}$ at low temperatures. The obtained values of $\chi_0$, $C_{LT}$ and $\theta_{LT}$ are ~ ($2.67 \times 10^{-3} \pm 0.096 \times 10^{-3}$) emu/mole-Oe, ~ ($0.168 \pm 0.001$) emu/mol-Oe-K and ~ ($- 0.372 \pm 0.005$) K, respectively. The obtained $C_{LT}$ gives the effective moment equal to 1.16 $\mu_B$ which is very less than the theoretically calculated value (~ 2.54 $\mu_B$) for free $Ce^{3+}$ ion. This reduced value can be attributed to the presence of $J = ½$ doublet at $LT$ regime with ground state of $Ce^{3+}$ ion having $J_{eff} = ½$ and $g_J = 1.20$, obtained from the value of $C_{LT}$. The negative value of $\theta_{LT}$ signifies the presence of weak AFM correlations among the $Ce^{3+}$ spins. Furthermore, $H$ dependent magnetization ($M$) measurement at 2 K ($> \theta_{LT}$) is performed up to a field of 60 kOe and is shown in the Fig. 6. It displays linear behaviour up to 10 kOe and then starts showing the features of non-linearity at higher fields. The red solid line shows the fitting to the simulation of the Brillouin function for the $J = ½$ given by the equation:

$$\frac{M}{M_S} = \tanh\left(\frac{g_J J \mu_B H}{k_B T}\right) \quad \ldots\ldots(1)$$

where, $g_J$, $J$, $\mu_B$ and $k_B$ are Lande's g-factor, total angular momentum, Bohr magneton and Boltzmann constant, respectively. The best fit to the equation with $J = ½$ yields the value of $g_J \sim 1.13 \pm 0.02$, which matches very well with the one obtained from the $LT$ $\chi_{DC}^{-1}(T)$. Thus, our magnetic analysis shows the signature of presence of $J_{eff} = ½$ state at low temperatures.

## 3.3 Heat Capacity

To probe the low-$T$ thermodynamics of CeVO$_4$, $T$ dependent $C$ measurements are performed down to 1.8 K at various applied $H$ up to 70 kOe. The zero field $T$ dependent data is fitted with an equation consisting of linear combination of Debye and Einstein terms [46]:

$$C(T) = mC_{Debye}(T) + (1-m)\,C_{Einstein}(T) \quad \ldots\ldots\ldots (2)$$

where $m$ denotes the weightage of the Debye term. $C_{Debye}$ and $C_{Einstein}$ are the Debye and Einstein contributions to the lattice heat capacity, respectively, which are given by the equations mentioned below.

$$C_{v\,Debye}(T) = 9nR \left(\frac{T}{\theta_D}\right)^3 \int_0^{\frac{\theta_D}{T}} \frac{x^4 e^x}{(e^x - 1)^2} \quad \ldots\ldots (2\,(a))$$

$$C_{v\,Einstein}(T) = 3nR \left(\frac{\theta_E}{T}\right)^2 \frac{e^{\frac{\theta_E}{T}}}{(e^{\frac{\theta_E}{T}} - 1)^2} \quad \ldots\ldots (2\,(b))$$

where n, R, $\theta_D$ and $\theta_E$ are the number of atoms per formula unit, universal gas constant, Debye temperature and Einstein temperature, respectively. Fig. 7 depicts the temperature response of $C$ measured at 0 Oe, along with solid red line showing the fitting to equation (2). The obtained parameters from the fitting are $\theta_D \sim (348 \pm 3)$ K, $\theta_E \sim (939 \pm 19)$ K and m $\sim$ (0.83 $\pm$ 0.01). Additionally, inset (a) of Fig. 7 shows the same curve in a small temperature regime (150 – 160 K). No feature was noted in the curve (in contrast to the $\chi_{DC}$ curve at 100 Oe), implying the absence of any kind of magnetic ordering. Further, the inset (b) of Fig. 7 shows the expanded view of the same data below 15 K. An upturn is clearly visible around 5 K, as shown by an arrow. This upturn is attributed to emerge from the Schottky anomaly of Ce$^{3+}$ ions. This upturn exhibits field dependent behaviour and shows an increment on increasing the $H$ (Fig. 8 (a)). At 70 kOe, it exhibits a hump centred around 2.5 K, implying that this hump is possibly present at $T$ below 1.8 K for lower fields. This behaviour can be ascribed to the Zeeman splitting of the ground state Kramer doublet ($J = 1/2$) on application of $H$. In order to further understand it, we have used two-level Schottky scheme given by the equation [18]:

$$C(H\,\text{kOe}) - C(0\,\text{kOe}) = N_A k_B \frac{(\Delta/T)^2 e^{-\Delta/T}}{(1 + e^{-\Delta/T})^2} \quad \ldots\ldots (3)$$

where, $\Delta$ is the Zeeman energy gap between the energy levels of the ground state doublet. Fig. 8 (b) shows the data corresponding to $C(H\,\text{kOe}) - C(0\,\text{kOe})$. The solid red lines show the best fitting to the two-level Schottky scheme defined in the equation (3). The obtained value of $\varDelta$ follows a linear behaviour with respect to $H$, as clearly visible from the inset of Fig. 8 (b), implying that the splitting between the ground state doublet increases linearly with externally applied magnetic fields. This trend found in $\varDelta$ is fitted with an equation $\varDelta = g_J\mu_B H$, as shown by solid red line (inset of Fig. 8 (b)). The obtained value of $g_J$ from the slope of fit is found to be ~ (1.073 ± 0.055), which is in good agreement with our magnetic studies. Hence, it can be said that, this observation from the heat capacity, along with magnetic measurements confirm the presence of $J_{eff} = ½$ ground state at low temperature in CeVO$_4$. Thus, our studies suggest that CeVO$_4$ is a potential candidate to study the physics behind the $J_{eff} = ½$ in 3D spin systems.

## 4. Discussion

The realization of $J_{eff} = ½$ magnetic ground state in rare earth magnetic systems attract great attention, as it can be an alternative to Cu-based spin ($S$) = ½ systems. In the presence of strong spin orbit coupling (SOC) and crystal field effects, the rare earth ions with odd number of 4$f$ electrons can be treated as Kramer's doublet with effective spin, $J_{eff} = ½$. In this context, a number of Yb-based quantum magnets have shown experimental signatures of $J_{eff} = ½$ magnetic ground state [11 – 18]. The eight-fold degenerate $J = 7/2$ states of the Yb$^{3+}$ ions split into four Kramer's doublet due to combined effect of SOC and crystalline electric field. In these systems, the ground state and first excited state are separated by a large energy gap such that the low temperature magnetic ground state can be ascribed to a pseudospin – ½ with effective $g_J$ parameter. The compounds with Ce$^{3+}$ ions also exhibit odd number of 4$f$ electrons with ground state term symbol $^2F_{5/2}$. The six-fold degenerate $J = 5/2$ states split into three Kramer's doublets. Also, we have compared our experimental observations on CeVO$_4$ with other conventionally reported Yb-based systems. A good number of Yb-based chalcogenides, AYbX$_2$ (A = monovalent ions like Li, Na, Ag; X = divalent chalcogen ions like O, S, Se) are also reported to exhibit $J_{eff} = ½$ ground state [9 – 14]. Namely, NaYbS$_2$ [10], NaYbO$_2$ [11], and NaYbSe$_2$ [12] exhibits large energy gap between the ground state and first excited state which is around 200 K, 400 K and 180 K, respectively. This energy gap is directly related to the CW temperature obtained from the CW fitting in the high temperature PM regime. At low temperatures, considerably lower than this energy gap, the magnetic properties are well described by the effective spin – ½ state. In CeVO$_4$, the large value of the CW temperature (~ - 120 K) signifies the presence of large energy gap between the ground state and first excited state. The low temperature magnetic and heat capacity studies yield the signatures of magnetic ground state associated with $J_{eff} = ½$. As mentioned above, till now, such kind of study was mainly focussed on transition metal-based system having spin angular momentum (S = ½) and 2D triangular Yb-based systems. To the best of our knowledge, this kind of study is not reported in 3D spin systems and hence, we believe that our study might be beneficial to explore the possibility of effective spin – ½ state in other 3D spin systems.

## 5. Conclusion

In conclusion, we have performed structural, magnetic and thermodynamic measurements on rare earth orthovanadate, $CeVO_4$. $T$ dependent $\chi_{DC}$, AC susceptibility and $C$ studies at zero field reveal the absence of any kind of LRO and spin freezing down to 1.8 K. The effective moment extracted from the Curie constant (obtained from the low temperature $\chi_{DC}$ data) is consistent with $J_{eff}$ = ½. The $M$ ($H$) curve at 2 K ($> \theta_{LT}$) shows a good fit to Brillouin function for $J$ = ½ with $g_J \sim 1.13$, which matches very well with the value of $g_J$, obtained from $T$ dependent $\chi_{DC}^{-1}$ and $C$ data. $H$ dependent $C$ is found to be in good agreement with two-level Schottky scheme. All these experimental observations confirm the presence of $J_{eff}$ = ½ at low temperature. Further, low temperature microscopic probes like muon-spin resonance and inelastic neutron scattering, might prove to be more effective to explore the nature of ground state of $CeVO_4$.

**Acknowledgements**

The authors acknowledge IIT Mandi for the experimental facilities and financial support.

**Data availability statement**

All data that support the findings of this study are included within the article.

Table 1: Structural parameters obtained from the Rietveld refinement of the XRD data at 300 K. The $O_c$ and $O_e$ designations correspond to the compressed and elongated Ce-O bonds, respectively. The number given with × sign defines different bond angles in $CeO_8$ polyhedral and $VO_4$ tetrahedra.

| Atom | Wyckoff position | x | y | z | Occupancy |
|---|---|---|---|---|---|
| Ce | 4a | 0 | 0.75 | 0.125 | 1 |
| V | 4b | 0 | 0.25 | 0.375 | 1 |
| O | 16h | 0 | 0.0699 | 0.2191 | 4 |
| **Bond** | **Distance (Å)** | **Bond** | **Angle (º)** | **Bond** | **Angle (º)** |
| Ce-$O_c$ | 2.444 (7) | (O-Ce-O) × 2 | 61.596 (3) | (O-Ce-O) × 4 | 93.582 (4) |
| Ce-$O_e$ | 2.602 (7) | (O-Ce-O) × 4 | 73.678 (17) | (O-V-O) × 2 | 105.503 (3) |
| V-O | 1.673 (4) | (O-Ce-O) × 8 | 77.601 (7) | (O-V-O) × 4 | 111.491 (12) |

Table 2: Comparison of structural parameters of $CeVO_4$ with other members of the series $RVO_4$ [from Ref. 33].

| $RVO_4$ | Ce (This work) | Ce | Pr | Nd | Eu | Tb | Dy | Ho | Er | Tm | Yb | Lu |
|---|---|---|---|---|---|---|---|---|---|---|---|---|
| a (Å) | 7.399 | 7.400 | 7.363 | 7.331 | 7.186 | 7.177 | 7.146 | 7.122 | 7.096 | 7.068 | 7.043 | 7.025 |
| c (Å) | 6.496 | 6.497 | 6.465 | 6.435 | 6.331 | 6.326 | 6.307 | 6.289 | 6.273 | 6.259 | 6.247 | 6.234 |
| V (Å³) | 355.62 | 355.83 | 350.51 | 345.85 | 326.92 | 325.90 | 322.07 | 319.06 | 315.84 | 312.71 | 309.85 | 307.72 |

**Figures:**

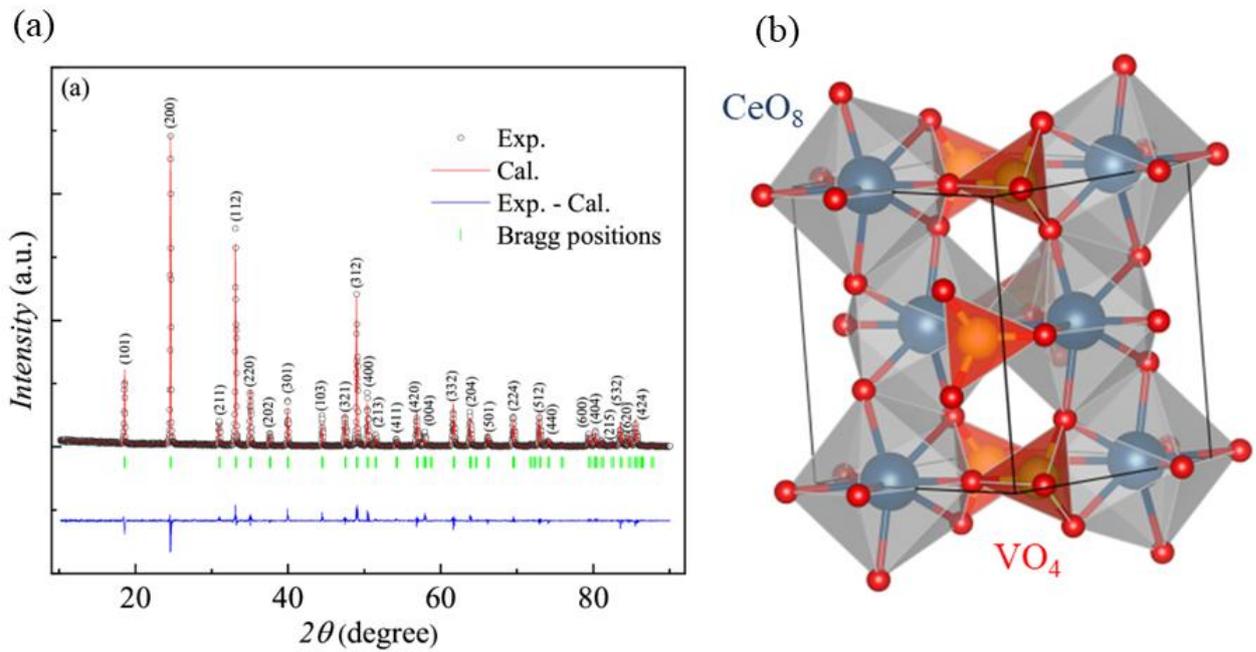

**Fig 1:** (a) Rietveld refined powder-XRD pattern of $CeVO_4$. The black open circles indicate the experimental data, while the Rietveld refinement fit is shown as solid red line. The difference curve and the Bragg positions are shown by solid blue line and green vertical lines, respectively. (b) Crystal structure of $CeVO_4$, comprised of $CeO_8$ polyhedral and $VO_4$ tetrahedra.

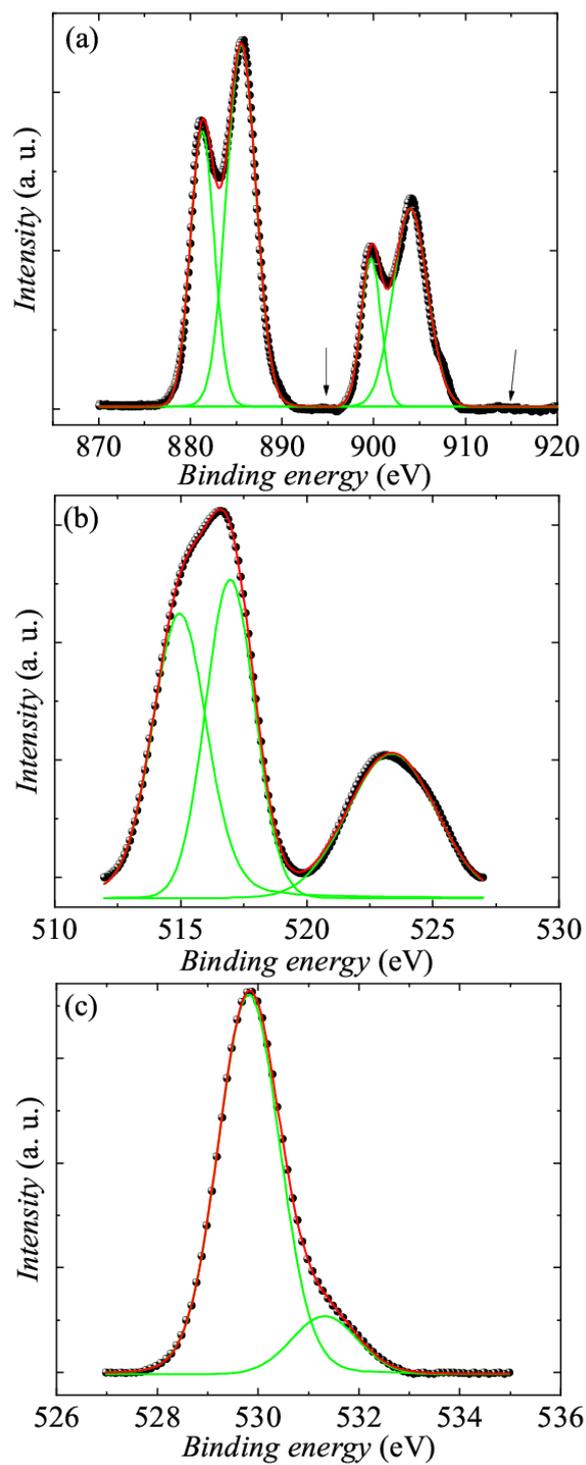

**Fig. 2:** (a, b, c) Room temperature X-ray photoemission spectra (XPS) of the core level of Ce 3*d*, V 2p, and O 1*s,* respectively. The solid red and green lines show the resultant fit and individual peak fittings, respectively.

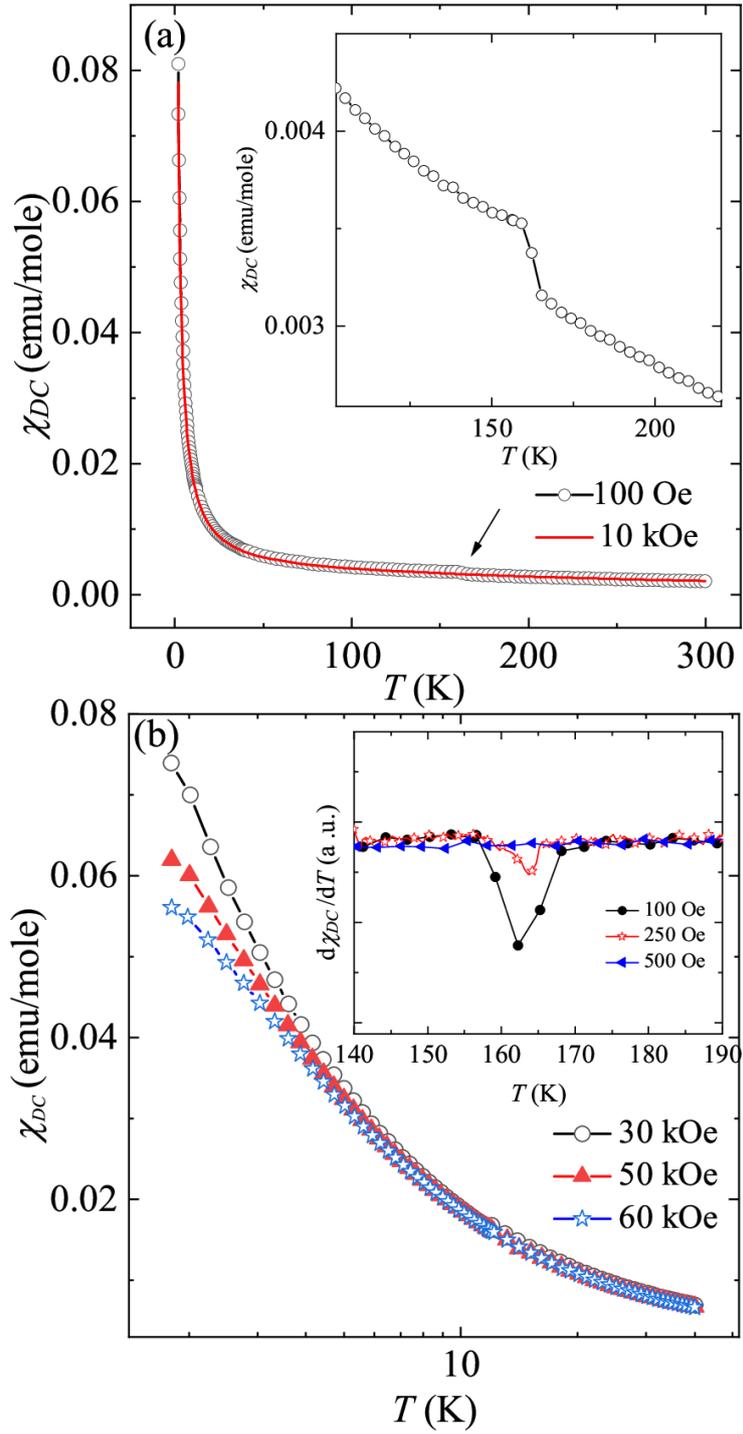

**Fig. 3:** (a) $T$ dependent $\chi_{DC}$ at 100 Oe and 10 kOe. The arrow indicates the inflection point observed in the 100 Oe curve. Inset: Same curve in the expanded region in the neighbourhood of inflection point. (b) $\chi_{DC}$ vs $T$ curves obtained at various applied fields on the log scale. Inset: Derivative of $\chi_{DC}$ with respect to temperature at various fields up to 500 Oe measured under ZFC protocol.

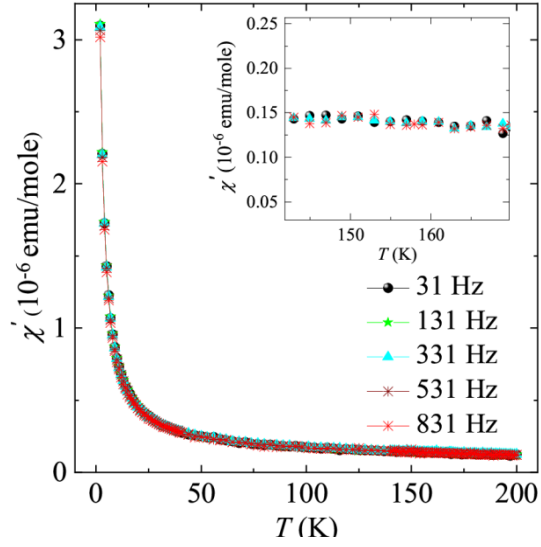

**Fig. 4:** *T* dependence of $\chi'$ at various frequencies 31 – 831 Hz for zero applied DC field. Inset shows the expanded view of the same data.

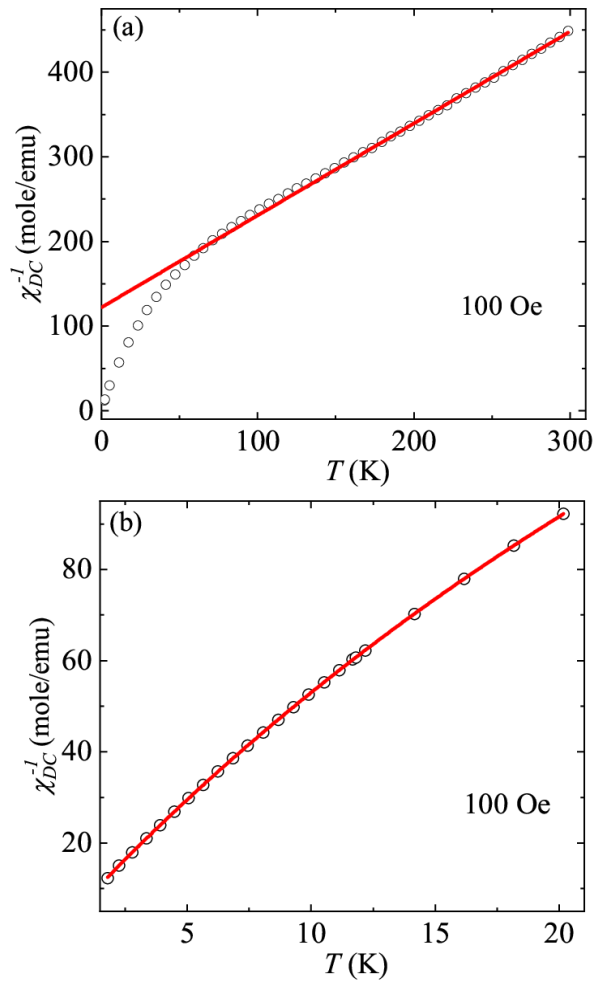

**Fig. 5:** (a) $\chi_{DC}^{-1}$ as a function of *T* at 100 Oe. The solid red line shows the fitting to the CW law in higher temperature region. (b) *T* dependent $\chi_{DC}^{-1}$ at 100 Oe below 20 K, with solid red line showing the fit to the modified CW law.

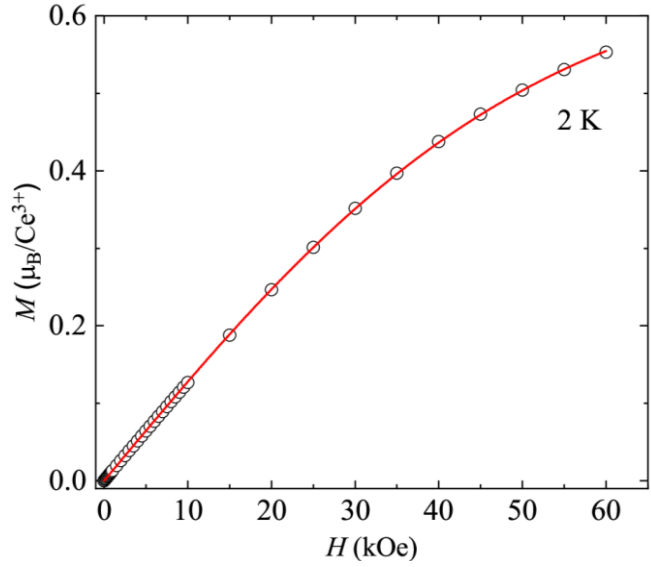

**Fig. 6:** Isothermal magnetization curve, *M* (*H*) at 2 K. The solid red line shows the simulation to the Brillouin function for *J* = ½, with $g_J$ ~ 1.13.

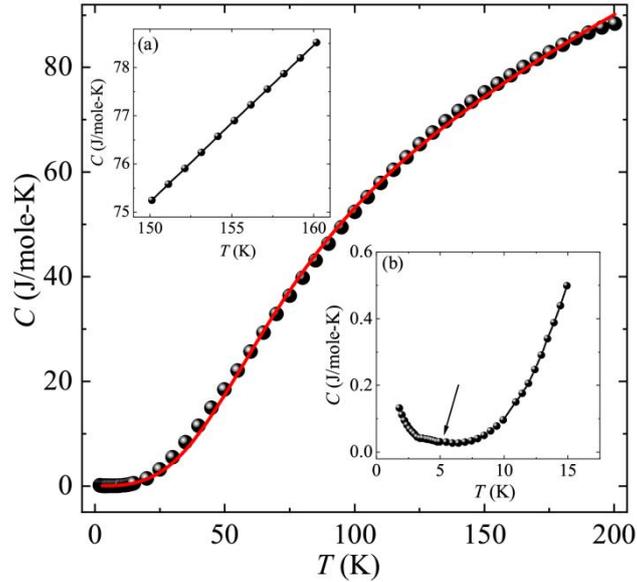

**Fig. 7:** *T* dependent heat capacity at 0 Oe up to 200 K with solid red line showing fit to the equation (2). Inset (a) shows the same data in the region where a step like behaviour is observed in $\chi_{DC}$. Inset (b) shows the expanded view of the same data below 15 K with the solid black arrow indicating the upturn.

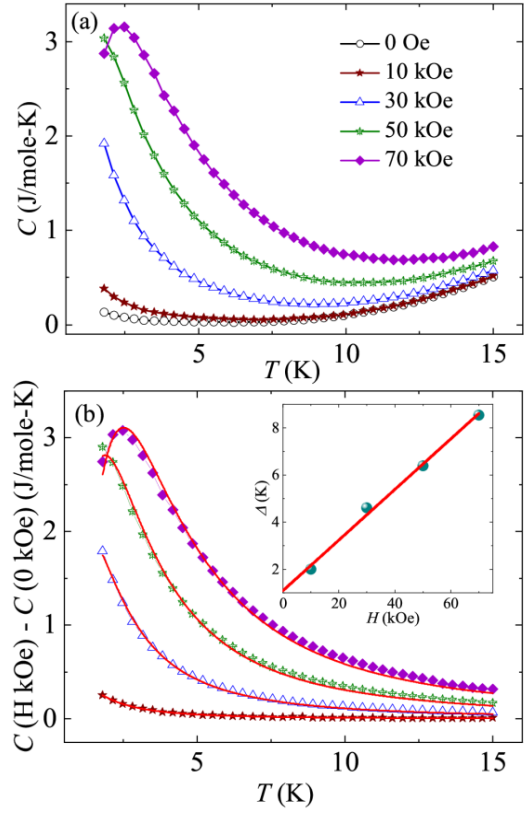

**Fig. 8:** (a) *T* response of heat capacity at various applied *H* up to 70 kOe. (b) Heat capacity after subtracting the zero-field data. The solid red lines show the fit to two-level Schottky scheme. Inset: Energy gap variation with the externally applied *H* along with the linear fitting (shown by the solid red line).